\documentclass[a4paper]{jpconf}

\usepackage{amsmath,amssymb,amsthm,eucal,mathrsfs,array,color,enumitem}
\usepackage[square,sort&compress,numbers,sort]{natbib}
\usepackage{graphicx}
\usepackage{tikz}
\usetikzlibrary{positioning}

\DeclareSymbolFont{largesymbols}{OMX}{zplm}{m}{n} 

\setitemize{leftmargin=*}                
\setenumerate{leftmargin=*,              
              label=\textup{(\arabic*)}} 

\newcolumntype{C}{>{$}c<{$}} 

\numberwithin{equation}{section}

\allowdisplaybreaks

\DeclareMathOperator{\vsp}{span}

\newcommand{\alg}[1]{\mathfrak{#1}} 
\newcommand{\grp}[1]{\mathsf{#1}} 

\newcommand{\func}[2]{#1 \left( #2 \right)} 
\newcommand{\tfunc}[2]{#1 \bigl( #2 \bigr)} 

\newcommand{\brac}[1]{\left( #1 \right)}

\newcommand{\set}[1]{\left\{ #1 \right\}}

\newcommand{\st}{\mspace{5mu} : \mspace{5mu}} 

\newcommand{\ZZ}{\mathbb{Z}}

\newcommand{\RR}{\mathbb{R}}

\newcommand{\pd}{\partial}     
\newcommand{\ahol}[1]{\overline{#1}}

\newcommand{\dd}{\mathrm{d}}   
\newcommand{\ii}{\mathfrak{i}} 
\newcommand{\ee}{\mathsf{e}}   

\newcommand{\wun}{\mathbf{1}}  

\newcommand{\normord}[1]{\mbox{${} : #1 : {}$}} 

\newcommand{\Res}[1]{#1{}\hspace{-0.2em} \downarrow {}}
\newcommand{\Ind}[1]{#1{}\hspace{-0.2em} \uparrow {}}

\newcommand{\Ra}{\Rightarrow}
\newcommand{\ra}{\rightarrow}
\newcommand{\lra}{\longrightarrow}
\newcommand{\ses}[3]{0 \ra #1 \ra #2 \ra #3 \ra 0}                                  
\newcommand{\dses}[5]{0 \lra #1 \overset{#2}{\lra} #3 \overset{#4}{\lra} #5 \lra 0} 
\newcommand{\res}[4]{\cdots \lra #4 \lra #3 \lra #2 \lra #1 \lra 0}                 

\newcommand{\affine}[1]{\widehat{#1}}
\newcommand{\SLG}[2]{\grp{#1} \bigl( #2 \bigr)}                             
\newcommand{\SLSA}[3]{\alg{#1} \left( #2 \middle\vert #3 \right)}           
\newcommand{\AKMA}[2]{\affine{\alg{#1}} \left( #2 \right)}                  
\newcommand{\AKMSA}[3]{\affine{\alg{#1}} \left( #2 \middle\vert #3 \right)} 

\newcommand{\VOA}[1]{\mathsf{#1}}                                           

\newcommand{\sfaut}{\sigma}                           
\newcommand{\sfmod}[2]{\tfunc{\sfaut^{#1}}{#2}}       

\newcommand{\Vac}{\mathcal{V}}             
\newcommand{\Para}[1]{\mathcal{W}_{#1}}    
\newcommand{\Fock}[1]{\mathcal{F}_{#1}}    
\newcommand{\SingIrr}[1]{\mathcal{L}_{#1}} 
\newcommand{\TripIrr}[1]{\mathbb{L}_{#1}}  
\newcommand{\TripSt}[1]{\mathbb{A}_{#1}}   

\newcommand{\SC}{\mathcal{J}}               

\newcommand{\traceover}[1]{\tr_{\raisebox{-3pt}{$\scriptstyle #1$}}}
\newcommand{\chmap}{\mathrm{ch}}
\newcommand{\torus}{\mathrm{tor}}
\newcommand{\Gr}[1]{\bigl[ #1 \bigr]}           
\newcommand{\ch}[1]{\chmap \Gr{#1}}             
\newcommand{\fch}[2]{\tfunc{\ch{#1}}{#2}}       

\newcommand{\jth}[1]{\vartheta_{#1}}            

\newcommand{\modS}{\mathsf{S}}                        
\newcommand{\modT}{\mathsf{T}}                        
\newcommand{\modA}{\mathsf{A}}                        

\newcommand{\Smat}[2]{\modS \bigl[ #1 \ra #2 \bigr]}  

\newcommand{\fuse}{\mathbin{\times}}                              
\newcommand{\fuscoeff}[3]{\genfrac{[}{]}{0pt}{0}{#3}{#1 \ \ #2}}  

\newcommand{\eqnref}[1]{Equation~\eqref{#1}}

\newcommand{\secref}[1]{Section~\ref{#1}}

\newcommand{\exref}[1]{Example~\ref{#1}}

\newcommand{\cft}{conformal field theory}
\newcommand{\cfts}{conformal field theories}

\newcommand{\lcft}{logarithmic conformal field theory}
\newcommand{\lcfts}{logarithmic conformal field theories}
\newcommand{\WZW}{Wess-Zumino-Witten}

\newcommand{\opes}{operator product expansions}
\newcommand{\hws}{highest weight state}
\newcommand{\hwss}{highest weight states}

\newcommand{\hwms}{highest weight modules}
\newcommand{\voa}{vertex operator algebra}
\newcommand{\voas}{vertex operator algebras}

\theoremstyle{definition}
\newtheorem{ex}{Example}

\begin{document}

\title{The Verlinde formula in logarithmic CFT}

\author{David Ridout$^{1,2}$ and Simon Wood$^{1,2}$}

\address{$^1$ Department of Theoretical Physics, Research School of Physics and Engineering, Australian National University, Acton, ACT 2600, Australia}
\address{$^2$ Mathematical Sciences Institute, Australian National University, Acton, ACT 2600, Australia}

\ead{david.ridout@anu.edu.au, simon.wood@anu.edu.au}

\begin{abstract}
In rational conformal field theory, the Verlinde formula computes the fusion coefficients from the modular S-transformations of the characters of the chiral algebra's representations.  Generalising this formula to logarithmic models has proven rather difficult for a variety of reasons.  Here, a recently proposed formalism \cite{CreLog13} for the modular properties of certain classes of logarithmic theories is reviewed, and refined, using simple examples.  A formalism addressing fusion rules in simple current extensions is also reviewed as a means to tackle logarithmic theories to which the proposed modular formalism does not directly apply.
\end{abstract}

\section{Introduction} \label{sec:Intro}

Logarithmic \cft{} typically refers to a two-dimensional conformally invariant quantum field theory for which there exist correlation functions exhibiting logarithmic singularities.  The first study of a \lcft{} appears to be that of Rozansky and Saleur \cite{RozQua92} where the model under consideration possesses symmetries that include the affine Kac-Moody superalgebra $\AKMSA{gl}{1}{1}$ of central charge $c=0$.  Shortly thereafter, Gurarie noted \cite{GurLog93} the appearance of logarithmic singularities in the $c=-2$ fermionic ghost system and showed that a sufficient condition for their presence was the non-diagonalisability of the hamiltonian $\mathbf{H} = L_0 + \ahol{L}_0$.  (He also seems to be responsible for the appellation ``\lcft{}''.)  We refer to \cite{RidLog13} for a recent collection of reviews on this topic.

As with rational \cft{}, much of the analysis of logarithmic theories begins with identifying the physically relevant spectrum of representations of the chiral symmetry algebra (\voa{}).  Unlike the rational case, where the representations are all completely reducible, the spectrum of a logarithmic theory necessarily includes representations which are reducible, yet indecomposable.\footnote{We recall that a representation is said to be indecomposable if it cannot be written as the direct sum of two non-trivial representations.}  In particular, the eigenvectors of $\mathbf{H}$ always span a subrepresentation which will be proper when $\mathbf{H}$ has non-trivial Jordan blocks.  Whilst the failure of complete reducibility does not by itself guarantee Jordan blocks, hence logarithmic singularities, it seems that all the known \cfts{} whose spectrum includes reducible but indecomposable representations are logarithmic.  Moreover, being logarithmic seems to be a generic property for \cfts{}, rather than a pathological one.

In both the rational and logarithmic cases, the issue of deciding whether a given spectrum of \voa{} representations is physically relevant or not deserves comment.  In any physical application, this would hopefully be clear.  We will, however, also use the term ``physically relevant'' in an abstract setting to describe mathematical objects or structures that satisfy certain conditions that one would expect to hold from a field-theoretic perspective.  In particular, a physically relevant spectrum of representations (category of modules) must be closed under conjugation, to guarantee non-degenerate two-point functions, and fusion, to guarantee closure under \opes{}.  Moreover, one must be able to form modular invariant partition functions from the characters.\footnote{One should add other requirements, in particular that the four-point functions satisfy crossing symmetry.  However, this condition will not be important for what follows.}

It is important to note, however, that \lcfts{} are necessarily non-unitary.  This is not a concern for applications to statistical mechanics where the unitarity of a given lattice model is of little importance to the study of its phase transitions.  Examples of lattice models whose scaling limits are known, or are believed, to be described by \lcfts{} include:
\begin{itemize}
\item Critical dense polymers \cite{PeaSol07}, where the chiral symmetry is Virasoro (or symplectic fermions \cite{MorMod13}) with $c=-2$.
\item Critical percolation \cite{ReaAss07,RasFus07,RidPer07,RidPer08}, where the symmetry is Virasoro with $c=0$.
\item Abelian sandpiles \cite{JenHei06,RueLog13}, where the symmetry is again Virasoro with $c=-2$, but the theory appears to differ from critical dense polymers.
\item The $\SLSA{gl}{1}{1}$ spin chain \cite{ReaAss07,GaiCon11,GaiLog13}, where the symmetry is symplectic fermions with $c=-2$.
\end{itemize}
Non-unitarity is, of course, a concern for string-theoretic applications, but here we note that it is not the underlying \cft{}, but only what survives the BRST quantisation (gauge-fixing) that needs to be unitary.  Logarithmic theories relevant to string theory include:
\begin{itemize}
\item Fermionic ($bc$) ghosts or symplectic fermions \cite{KauCur95,KauSym00}, where the chiral symmetry algebra is the affine Kac-Moody superalgebra $\AKMSA{psl}{1}{1}$ with $c=-2$.
\item Bosonic ($\beta\gamma$) ghosts of central charge $c=-1$ \cite{LesLog04,RidFus10} and $c=2$ \cite{RidBos14}.
\item \WZW{} models on Lie supergroups and their coset superspaces \cite{RozQua92,SalGL106,GotWZN07,CreRel11,QueSup13}.
\end{itemize}
There are also \lcfts{} that have been extensively studied for their own sake, in particular the $(1,p)$ triplet models \cite{GabRat96,GabLoc99,FeiMod06,GabFro08,AdaTri08,PeaInt08,NagTri11,TsuTen12} (see \cite{GaiLat12} for a proposed lattice-theoretic realisation), their $(p',p)$ generalisations \cite{FeiLog06,RasWEx09,GabFus09,WooFus10,AdaWAl09,GabMod11,AdaExp12,TsuExt13}, also known as the (W-extended) logarithmic minimal models, and the fractional level \WZW{} models with $\AKMA{sl}{2}$ \cite{GabFus01,LesLog04,RidSL208,RidFus10} and $\AKMA{sl}{3}$ \cite{AdaRea14} symmetry.  Recently, these models have been extended to $\AKMA{sl}{2}$ triplets \cite{SemLog13}.

The mathematical formalism of rational \cft{} is very rich and it is natural to try to extend the key insights to the logarithmic case.  In particular, the determination of the fusion rules in rational theories is often significantly simplified through the use of the Verlinde formula \cite{VerFus88}.  This relies on computing the modular S-transformations of the characters of the irreducible representations assuming, of course, that amenable character formulae can be found.  Unfortunately, it has been known for quite some time now \cite{FloMod96} that the irreducible characters of a \lcft{} need not carry a representation of the modular group, an observation which appears to completely invalidate the application of any Verlinde-type formula.

An example illustrating this dilemma is the $(1,2)$ triplet model \cite{GabRat96,GabLoc99} whose central charge is $c=-2$.  In this model, there are four irreducible representations, $\TripIrr{0}$, $\TripIrr{1}$, $\TripIrr{-1/8}$ and $\TripIrr{3/8}$, labelled by the conformal dimensions of their \hwss{}.  The corresponding characters are
\begin{equation} \label{ch:Trip}
\begin{aligned}
\fch{\TripIrr{0}}{\tau} &= \frac{1}{2} \brac{\frac{\func{\jth{1,2}}{0 ; \tau}}{\func{\eta}{\tau}} + \func{\eta}{\tau}^2}, \\
\fch{\TripIrr{1}}{\tau} &= \frac{1}{2} \brac{\frac{\func{\jth{1,2}}{0 ; \tau}}{\func{\eta}{\tau}} - \func{\eta}{\tau}^2},
\end{aligned}
\qquad
\begin{aligned}
\fch{\TripIrr{-1/8}}{\tau} &= \frac{\func{\jth{0,2}}{0 ; \tau}}{\func{\eta}{\tau}}, \\
\fch{\TripIrr{3/8}}{\tau} &= \frac{\func{\jth{2,2}}{0 ; \tau}}{\func{\eta}{\tau}}.
\end{aligned}
\end{equation}
One quickly checks that it is the factors of $\func{\eta}{\tau}^2$ that lead to $\tau$-dependent coefficients and spoil the modularity.  For example,
\begin{equation} \label{eq:TripNotMod}
\fch{\TripIrr{0}}{-1/\tau} = -\frac{\ii \tau}{2} \brac{\fch{\TripIrr{0}}{\tau} - \fch{\TripIrr{1}}{\tau}} + \frac{1}{4} \brac{\fch{\TripIrr{-1/8}}{\tau} - \fch{\TripIrr{3/8}}{\tau}}.
\end{equation}
This can be alleviated by extending the space of characters to include a new object
\begin{equation} \label{eq:ToralAmp}
\func{\torus}{\tau} = -\ii \tau \brac{\fch{\TripIrr{0}}{\tau} - \fch{\TripIrr{1}}{\tau}}
\end{equation}
and checking that the modular S- and T-transformations preserve the space spanned by the irreducible characters and $\torus$.

Unsurprisingly, there is a fair bit of literature addressing the issue of modularity for \lcfts{}.  In particular, Miyamoto showed \cite{MiyMod04} that one is only guaranteed to have an action of the modular group on the space of torus amplitudes (one-point functions on the torus).  Characters are examples of these amplitudes, as is the function $\torus$ of \eqref{eq:ToralAmp}, so Miyamoto's result meshes nicely with the extended character space of the $(1,2)$ triplet model.  However, the proof requires generalising the notion of torus amplitude (and hence character) to insert arbitrary elements of the chiral symmetry algebra into the trace.  While no additional elements were needed in the case of the $(1,2)$ triplet model, it is not clear whether this will be true more generally.  If not, then evaluating these refined characters and torus amplitudes leads to severe computational difficulties in all but the simplest examples.  It also leaves one wondering how such a result helps in determining modular invariant bulk partition functions (which are, after all, sesquilinear combinations of characters).

Finding modular invariants therefore seems to require new insight \cite{FucFro13}, but one can instead ask a different question.  Given a \lcft{}, is there a variant of the Verlinde formula which can be used to compute the fusion rules efficiently?  The technical answer to this question is ``no'', as long as we assume that the input to the Verlinde formula is the S-transformation of characters, because characters cannot distinguish between a representation and the direct sum of its irreducible constituents (composition factors).  Instead, the best that we can hope for is a Verlinde-type formula that determines the \emph{Grothendieck} fusion rules that are obtained from the genuine fusion rules by identifying representations with the same character.\footnote{There is a subtle mathematical question that we are overlooking here of whether the notion of Grothendieck fusion is well-defined.  See \cite{GabFus09} for an example where it is not.}

Is there a Verlinde formula that computes the Grothendieck fusion rules?  For the $(1,p)$ triplet models, the answer is ``yes'' \cite{FucNon04}.  Here, one computes the modular S-matrix $\modS$ on the basis of irreducible characters directly, so that the entries are $\tau$-dependent in general.  But then, the action is twisted by a carefully chosen $\tau$-dependent automorphy factor $\modA$ such that the product $\modS' = \modA \modS$ is both independent of $\tau$ and, with $\modT$, defines a representation of the modular group on character space.  Remarkably, this matrix $\modS'$ successfully block-diagonalises the Grothendieck fusion rules in a Verlinde-type formula given in \cite[p717]{FucNon04}.  Most importantly, the entries of the block-diagonal matrix are also expressed in terms of the entries of $\modS'$.

Admittedly, this formula is rather more complicated than the Verlinde formula of rational \cft{}.  However, this is to be expected from the non-diagonalisability of the fusion matrices in the logarithmic case.  But, there seems to be little understanding of why twisting by this automorphy factor $\modA$ is required, whether there are other possibilities for $\modA$, and whether, for more general models, the existence of such a factor is always guaranteed.  As far as we are aware, this type of Verlinde formula has not been generalised beyond the $(1,p)$ triplet models.  Moreover, there are now several other proposals \cite{FloVer07,GabFro08,GaiRad09,PeaGro10} addressing seemingly different Verlinde formulae for these theories.

Rather than try to compare and contrast these proposals, which would be a worthy goal in itself, we shall instead turn to a rather different formalism which has recently been introduced \cite{CreLog13} for modular properties and Verlinde formulae in a more general class of \lcfts{}.  One difference between this proposal and those mentioned above is that it does not directly apply to the triplet models.  However, this is not a serious restriction because the triplet models have close relatives to which the new formalism does apply.  We shall explain how this allows us to compute Grothendieck fusion rules for the triplet models in \secref{sec:SimpCurrExt}.

\section{Modularity for logarithmic CFT} \label{sec:ModLCFT}

The formalism discussed in this section arises from the observation \cite{CreRel11,CreLog13} that the majority of the \lcfts{} that have been adequately analysed to date all share certain key features.  Moreover, the known exceptions may all be understood as simple current extensions of theories with the desired features.  We will therefore describe these features before explaining, with the aid of examples, how they allow one to straightforwardly compute modular transformations and apply the Verlinde formula to obtain Grothendieck fusion coefficients.

The formalism centres \cite{CreLog13} about the identification of a collection of representations of the chiral algebra (\voa{}) that we call the \emph{standard modules}.  These standard modules and their characters should have the following properties:
\begin{enumerate}
\item The characters of the standard modules are parametrised by a measurable space $(M,\mu)$. \label{it:meas}
\item The irreducible standard modules are called \emph{typical}.  The complement of the subset of $M$ corresponding to the typical characters has measure zero with respect to $\mu$. \label{it:typ}
\item The only indecomposable module in the spectrum having a typical module (irreducible standard) as a composition factor (irreducible subquotient) is that typical module.\footnote{Another way of stating this would be to say that typical modules are both projective and injective in the physically relevant category of modules.  However, this is extremely difficult to check in practice.} \label{it:proj}
\item The characters of the standard modules form a (topological) basis for the space of all characters.  In particular, the standard module characters are linearly independent. \label{it:res}
\item The characters of the standard modules span a representation of the modular group $\SLG{SL}{2;\ZZ}$.  In the standard basis, the S-transformation defines a symmetric, unitary operator whose square may be identified with conjugation. \label{it:S}
\item Combining the S-transformation with the standard Verlinde formula leads to the Grothendieck fusion rules (with non-negative integer coefficients). \label{it:Ver}
\end{enumerate}
Any module in the spectrum that is not a direct sum of typical modules is called \emph{atypical}, the nomenclature here being inspired by that of Kac for Lie superalgebra representations.  We remark that there will usually be more than one inequivalent standard module for each atypical parameter $m \in M$, but each of these atypical standard modules will have the same character.

We will not give any general means for identifying the standard modules in a given theory beyond the obvious need to satisfy the above requirements; in examples, this identification is often clear.  In this vein, we mention that verifying \ref{it:proj} rigorously is almost always too difficult at present, so we usually have to content ourselves with conjecturing that it holds.  We also remark that \ref{it:res} is generally seen to be satisfied by the existence of resolutions for the irreducible atypical modules in terms of atypical standard modules.  The Euler-Poincar\'{e} principle then expresses the character of each irreducible as an (infinite) alternating sum of standard characters.

The main observation that we want to emphasise is that the typical modules behave, for all intents and purposes, as if they were in a rational \cft{}.  Indeed, rational theories fit into the above formalism as the special case in which all the standard modules are typical.  The logarithmic nature of the theory is therefore concentrated in the atypical modules and these are required to form a set of measure zero in character-space.  This means that, heuristically at least, any potential problems that the logarithmic structure would create are measure-theoretically void.  While there is an open question as to how one identifies the appropriate measure in general, we will see that in examples, the natural choice works.

The simplest (non-rational) example illustrating this formalism is the (uncompactified) free boson.  However, as this model is not logarithmic, we will refer the reader to \cite[Sec.~1.2]{CreLog13} and instead turn to a closely related \lcft{}.
\begin{ex}[The singlet model \protect{\cite{CreLog13,RidMod13}}] \label{ex:Sing}
The fermionic $bc$ ghost system at $c=-2$ contains the symplectic fermions algebra as the subalgebra generated by $b$ and $\pd c$.  This subalgebra in turn contains a subalgebra corresponding to the states of zero ghost number.  This is the $(1,2)$ singlet algebra and it is generated by the energy-momentum tensor and a single Virasoro primary $W$ of conformal dimension $3$.
\begin{enumerate}
\item The singlet algebra possesses a continuous family of \hwms{} $\Fock{\lambda}$, with $\lambda \in \RR$.  These are the standard modules.  Their characters count the eigenvalues of $L_0$, but these do not distinguish conjugate modules.  We therefore augment them by counting the eigenvalues of a shifted ghost number, the zero mode of $G = \normord{bc} - \tfrac{1}{2}$, and the central element $\wun$:
\begin{equation} \label{ch:StandSing}
\fch{\Fock{\lambda}}{y;z;q} = \traceover{\Fock{\lambda}} y^{\wun} z^{G_0} q^{L_0 + 1/12} = \frac{y z^{\lambda-1/2} q^{\brac{\lambda-1/2}^2 / 2}}{\func{\eta}{q}}.
\end{equation}
\item The standard modules are irreducible if and only if $\lambda \notin \ZZ$.  We therefore equip $\RR$ with Lebesgue measure $\mu$, noting that $\func{\mu}{\ZZ} = 0$.
\item It has not yet been proved that the typicals are projective and injective.  However, this is a natural conjecture that passes non-trivial consistency checks such as the behaviour of typicals under fusion.
\item The standard characters \eqref{ch:StandSing} are clearly linearly independent.  The atypical standard modules $\Fock{\lambda}$, with $\lambda \in \ZZ$, are built from two composition factors (irreducible subquotients).  The structure is given by the following short exact sequences:\footnote{Exactness means that at each position $\cdots \overset{f}{\lra} M \overset{g}{\lra} \cdots$ in the sequence, the image of $f$ coincides with the kernel of $g$.  If $\ses{L}{M}{N}$ is exact, then $L$ is a submodule of $M$ and $M / L$ is isomorphic to $N$.}
\begin{equation} \label{es:Sing}
\dses{\SingIrr{\lambda}}{}{\Fock{\lambda}}{}{\SingIrr{\lambda-1}}.
\end{equation}
We then splice (see \cite[App.~A.2]{CreLog13} for example) these exact sequences to obtain resolutions:
\begin{equation}
\res{\SingIrr{\lambda}}{\Fock{\lambda+1}}{\Fock{\lambda+2}}{\Fock{\lambda+3}}.
\end{equation}
This is also an exact sequence, so the Euler-Poincar\'{e} principle leads to the character formula
\begin{equation} \label{ch:AtypSing}
\ch{\SingIrr{\lambda}} = \sum_{n=1}^{\infty} \brac{-1}^{n-1} \ch{\Fock{\lambda+n}} \qquad \text{(\(\lambda \in \ZZ\)),}
\end{equation}
verifying that the standard characters form a topological basis of the space of all characters.
\item The standard singlet characters span a representation of the modular group whose dimension is uncountably-infinite.  The S-transformation is essentially a Fourier transform:
\begin{equation}
\begin{gathered}
\fch{\Fock{\lambda}}{\theta - \frac{\zeta^2}{2 \tau}; \frac{\zeta}{\tau}; -\frac{1}{\tau}} = \int_{\RR} \Smat{\Fock{\lambda}}{\Fock{\lambda'}} \fch{\Fock{\lambda'}}{\theta; \zeta; \tau} \: \dd \lambda', \\
\Smat{\Fock{\lambda}}{\Fock{\lambda'}} = \ee^{-2 \pi \ii \brac{\lambda - 1/2} \brac{\lambda' - 1/2}}.
\end{gathered}
\end{equation}
This is symmetric and unitary and it squares to the conjugation permutation $\lambda \leftrightarrow 1-\lambda$:
\begin{subequations}
\begin{gather}
\Smat{\Fock{\lambda}}{\Fock{\lambda'}} = \Smat{\Fock{\lambda'}}{\Fock{\lambda}}, \\
\int_{\RR} \Smat{\Fock{\lambda}}{\Fock{\lambda'}} \Smat{\Fock{\lambda''}}{\Fock{\lambda'}}^* \: \dd \lambda' = \func{\delta}{\lambda - \lambda''}, \\
\int_{\RR} \Smat{\Fock{\lambda}}{\Fock{\lambda'}} \Smat{\Fock{\lambda'}}{\Fock{\lambda''}} \: \dd \lambda' = \func{\delta}{\lambda + \lambda'' - 1}.
\end{gather}
\end{subequations}
The atypical S-transformations are now obtained from the character formula \eqref{ch:AtypSing}:
\begin{equation} \label{eq:AtypSSing}
\Smat{\SingIrr{\lambda}}{\Fock{\lambda'}} = \sum_{n=1}^{\infty} \brac{-1}^{n-1}\Smat{\Fock{\lambda + n}}{\Fock{\lambda'}}  = \frac{\ee^{-2 \pi \ii \lambda \brac{\lambda' - 1/2}}}{2 \cos \brac{\pi \brac{\lambda' - 1/2}}}.
\end{equation}
We note that the these elements \emph{diverge} precisely at the atypical parameters $\lambda' \in \ZZ$.
\item The (Grothendieck) fusion rules are given by the following version of the Verlinde formula:
\begin{equation} \label{eq:VerFB}
\begin{gathered}
\fuscoeff{\Fock{\lambda}}{\Fock{\lambda'}}{\Fock{\lambda''}} = \int_{\RR} \frac{\Smat{\Fock{\lambda}}{\Fock{\lambda'''}} \Smat{\Fock{\lambda'}}{\Fock{\lambda'''}} \Smat{\Fock{\lambda''}}{\Fock{\lambda'''}}^*}{\Smat{\SingIrr{0}}{\Fock{\lambda'''}}} \: \dd \lambda''', \\
\ch{\Fock{\lambda}} \fuse \ch{\Fock{\lambda'}} = \int_{\RR} \fuscoeff{\Fock{\lambda}}{\Fock{\lambda'}}{\Fock{\lambda''}} \ch{\Fock{\lambda''}} \: \dd \lambda''.
\end{gathered}
\end{equation}
Note that $\SingIrr{0}$ is the vacuum module.  There are atypical analogues obtained by replacing $\Fock{\lambda}$ or $\Fock{\lambda'}$ (but never $\Fock{\lambda''}$) by atypical modules.  We thereby obtain the Grothendieck fusion rules
\begin{equation}
\begin{gathered}
\ch{\SingIrr{\lambda}} \fuse \ch{\SingIrr{\lambda'}} = \ch{\SingIrr{\lambda + \lambda'}}, \qquad
\ch{\SingIrr{\lambda}} \fuse \ch{\Fock{\lambda'}} = \ch{\Fock{\lambda + \lambda'}}, \\
\ch{\Fock{\lambda}} \fuse \ch{\Fock{\lambda'}} = \ch{\Fock{\lambda + \lambda'}} + \ch{\Fock{\lambda + \lambda'-1}},
\end{gathered}
\end{equation}
confirming that the Grothendieck fusion coefficients are all non-negative integers.  The atypical-atypical rule lifts to a genuine fusion rule because the result is an irreducible character that determines the module uniquely.  The same is true for the atypical-typical rule because $\lambda + \lambda' \notin \ZZ$.  The typical-typical rule, however, can only be lifted to a direct sum when $\lambda + \lambda'$ is typical.  When $\lambda + \lambda' \in \ZZ$, we expect that the fusion product is a staggered module in the sense of \cite{RidSta09,CreLog13}.\footnote{Unfortunately, nobody seems to have computed fusion products for the $(1,2)$ singlet algebra.  However, this expectation is consistent with the known fusion rules for the $(1,2)$ \emph{triplet} algebra \cite{GabRat96} as we shall see.}
\end{enumerate}
\end{ex}

The observed divergence \eqref{eq:AtypSSing} of the atypical S-transformation kernel at atypical parameters seems to be a general feature of logarithmic theories.  For this reason, it is absolutely vital that the atypical standards correspond to a set of measure zero in parameter space.  We will develop a method to deal with the case when the atypical parameters have positive measure in \secref{sec:SimpCurrExt}.  Before turning to this, however, we detail another, slightly more involved, logarithmic example.
\begin{ex}[Bosonic ghosts \protect{\cite{RidBos14}}] \label{ex:Ghost}
The $\beta\gamma$ ghost system with $c=2$ possesses a unique highest weight representation, the vacuum module $\Vac$.  This module is irreducible.  However, the ghost algebra admits a family of spectral flow automorphisms $\sfaut^{\ell}$ which yields a family of mutually inequivalent representations $\sfmod{\ell}{\Vac}$, $\ell \in \ZZ$.  These are \emph{not} the standard modules of the theory because their characters do \emph{not} span a representation of the modular group.
\begin{enumerate}
\item The ghost algebra also admits a continuous family of \emph{parabolic} Verma modules $\Para{j}$, with $0<j<1$.  Moreover, there exist two additional inequivalent indecomposable modules, $\Para{0}^+$ and $\Para{0}^-$, characterised by the short exact sequences
\begin{equation} \label{es:Ghost}
\dses{\Vac}{}{\Para{0}^+}{}{\sfmod{-1}{\Vac}}, \qquad \dses{\sfmod{-1}{\Vac}}{}{\Para{0}^-}{}{\Vac}.
\end{equation}
The spectral flow images $\sfmod{\ell}{\Para{j}}$, together with the $\sfmod{\ell}{\Para{0}^{\pm}}$, are the standard modules.  The standard characters again involve the ghost number, obtained from the zero mode of $J = \normord{\beta\gamma}$, as well as the central element $\wun$:
\begin{equation}
\fch{\sfmod{\ell}{\Para{j}}}{y;z;q} = \traceover{\sfaut^{\ell}(\Para{j})} y^{\wun} z^{J_0} q^{L_0 - 1/12} = \frac{y z^j q^{\ell j + \ell \brac{\ell - 1} / 2}}{\func{\eta}{q}^2} \sum_{n \in \ZZ} z^n q^{n \ell}.
\end{equation}
The common character of $\sfmod{\ell}{\Para{0}^+}$ and $\sfmod{\ell}{\Para{0}^-}$ is also given by this formula (with $j$ set to $0$).  The standard characters are thus naturally parametrised by $j \in \RR / \ZZ \cong \mathrm{S}^1$ and $\ell \in \ZZ$.  We equip this space with the product $\mu$ of the standard Haar measure on $\mathrm{S}^1$ and counting measure on $\ZZ$.
\item The $\sfmod{\ell}{\Para{j}}$, with $j \neq 0$, are irreducible.  These are the typical modules:  $\func{\mu}{\set{0} \times \ZZ} = 0$.
\item Again, we content ourselves with conjecturing that the typicals are projective and injective.
\item Splicing \eqref{es:Ghost} to obtain resolutions, we arrive at the atypical character formula
\begin{equation} \label{ch:AtypGhost}
\ch{\sfmod{\ell}{\Vac}} = \sum_{n=1}^{\infty} \brac{-1}^{n-1} \ch{\sfmod{\ell+n}{\Para{0}^+}},
\end{equation}
which confirms that the standard characters form a topological basis.
\item The modular transformations take a rather more complicated form for this model:
\begin{equation}
\begin{aligned}
\modS &\colon \left( \theta; \zeta; \tau \right) \lra \left( \theta + \frac{\zeta^2}{2 \tau} - \frac{\zeta}{2 \tau} + \frac{\zeta}{2} + \frac{1}{2 \pi} \brac{\arg \tau - \frac{\pi}{2}}; \frac{\zeta}{\tau}; -\frac{1}{\tau} \right), \\
\modT &\colon \left( \theta; \zeta; \tau \right) \lra \left( \theta + \frac{1}{12}; \zeta; \tau + 1 \right).
\end{aligned}
\end{equation}
Nevertheless, they determine a representation of $\SLG{SL}{2;\ZZ}$ on the span of the standard characters with S-transformation
\begin{equation}
\begin{gathered}
\fch{\sfmod{\ell}{\Para{j}}}{\modS \left( \theta; \zeta; \tau \right)} = \sum_{\ell' \in \ZZ} \int_{\RR / \ZZ} \Smat{\sfmod{\ell}{\Para{j}}}{\sfmod{\ell'}{\Para{j'}}} \fch{\sfmod{\ell'}{\Para{j'}}}{\theta; \zeta; \tau} \: \dd j', \\
\Smat{\sfmod{\ell}{\Para{j}}}{\sfmod{\ell'}{\Para{j'}}} = \brac{-1}^{\ell+\ell'} \ee^{-2 \pi \ii \brac{\ell j' + \ell' j}}.
\end{gathered}
\end{equation}
It is easy to check that $\modS$ is symmetric, unitary and squares to conjugation.  The atypical S-transformation kernel
\begin{equation} \label{eq:AtypSMatGhost}
\Smat{\sfmod{\ell}{\Vac}}{\sfmod{\ell'}{\Para{j'}}} = \brac{-1}^{\ell+\ell'+1} \frac{\ee^{-2 \pi \ii \brac{\ell + 1/2} j'}}{\ee^{\ii \pi j'} - \ee^{-\ii \pi j'}},
\end{equation}
again diverges at the atypical point $j'=0$.
\item The obvious Verlinde formula (with sums over $\ell \in \ZZ$ and integrals over $j \in \RR / \ZZ$) gives the following Grothendieck fusion rules:
\begin{subequations}
\begin{align}
\ch{\sfmod{\ell}{\Vac}} \fuse \ch{\sfmod{\ell'}{\Vac}} &= \ch{\sfmod{\ell+\ell'}{\Vac}}, \\
\ch{\sfmod{\ell}{\Vac}} \fuse \ch{\sfmod{\ell'}{\Para{j'}}} &= \ch{\sfmod{\ell+\ell'}{\Para{j'}}}, \\
\ch{\sfmod{\ell}{\Para{j}}} \fuse \ch{\sfmod{\ell'}{\Para{j'}}} &= \ch{\sfmod{\ell+\ell'}{\Para{j + j'}}} + \ch{\sfmod{\ell+\ell'-1}{\Para{j + j'}}}.
\end{align}
\end{subequations}
The coefficients are clearly non-negative in each case.
\end{enumerate}
\end{ex}

We close this section by noting that this formalism has been extensively tested on other \lcfts{} including the fractional level $\AKMA{sl}{2}$ models \cite{CreMod12,CreMod13}, the $\AKMSA{gl}{1}{1}$ model \cite{CreRel11,CreWAl11} and its Takiff extension \cite{BabTak12}, the logarithmic minimal models \cite{MRR} and their $N=1$ extensions \cite{CRR}, and the $(p',p)$ singlet theories \cite{RidMod13}.  The Takiff model is of interest because it demonstrates that the formalism can naturally accommodate different degrees of atypicality.  The singlet theories with $p,p' \geq 2$ are interesting because it is known \cite{GabFus09} that their fusion product does not induce a well-defined product on the Grothendieck group because of the poor behaviour (non-rigidity) of the minimal model representations.  The formalism deals with this naturally, and rather effectively, by regarding the minimal model characters as being zero.\footnote{One is therefore led to wonder whether the minimal model representations should be excluded from the physically relevant category.  They do not seem to appear as direct summands in the bulk, boundary or defect sectors.  However, excluding them would amount to taking the physical category to be non-abelian.}

\section{Fusion Rules for Simple Current Extensions} \label{sec:SimpCurrExt}

We recall that a simple current may be generally defined as an indecomposable representation for which fusing with the simple current defines a permutation on the set of all indecomposables.  For the purposes of this article, we shall simplify this definition by replacing ``indecomposable'' by ``irreducible'', as in \cite{SchSim90,DonSim96}.  Examples of such simple currents then include the $\SingIrr{\lambda}$, with $\lambda \in \ZZ$, for the $(1,2)$ singlet model (\exref{ex:Sing}) and the $\sfmod{\ell}{\Vac}$, with $\ell \in \ZZ$, for the $c=2$ bosonic ghost system (\exref{ex:Ghost}).  In both cases, the simple currents are atypical irreducibles.

If the vacuum module is also irreducible, as is the case for both examples
considered in \secref{sec:ModLCFT}, then for every simple current $\SC$, there exists a simple current $\SC^{-1}$ such that their fusion gives the vacuum module, that is, $\SC$
has a fusion inverse.  Any given simple current (more generally, any set of simple currents) then generates a group $\grp{G}$ of simple currents under the fusion operation.  This group then generates an extended symmetry algebra in the sense that the representation $\bigoplus_{\SC \in \grp{G}} \SC$ carries the structure of a \voa{}.\footnote{There are some subtleties to this statement, however; see \cite{RidSU206,RidMin07,RidSL208} for example.}  Moreover, every representation $\mathcal{M}$ of the original symmetry algebra determines a representation of the extended symmetry algebra on $\bigoplus_{\SC \in \grp{G}} \brac{\SC \fuse \mathcal{M}}$.

Mathematically, this may be formalised using the notions of restriction and induction \cite{MiyRep98,LamInd01}:  Let $\VOA{A}$ denote the original \voa{}, $\VOA{B}$ its simple current extension by $\grp{G}$, and $\otimes_{\VOA{A}}$ and $\otimes_{\VOA{B}}$ their fusion products.  Then, restricting $\VOA{B}$ (interpreted as its vacuum module) to an $\VOA{A}$-module and inducing $\VOA{A}$ (also interpreted as its vacuum module) to a $\VOA{B}$-module amounts to
\begin{equation}
\Res{\VOA{B}} \cong \bigoplus_{\SC \in \grp{G}} \SC, \qquad
\Ind{\VOA{A}} = \VOA{B} \otimes_{\VOA{A}} \VOA{A} \cong \VOA{B}.
\end{equation}
Moreover, we may induce an $\VOA{A}$-module $\mathcal{M}$ to a $\VOA{B}$-module $\Ind{\mathcal{M}}$, obtaining
\begin{equation} \label{eq:IndMod}
\Ind{\mathcal{M}} = \VOA{B} \otimes_{\VOA{A}} \mathcal{M} \qquad \Ra \qquad
\Res{\brac{\Ind{\mathcal{M}}}} \cong \bigoplus_{\SC \in \grp{G}} \brac{\SC \otimes_{\VOA{A}} \mathcal{M}},
\end{equation}
as before.  This formalisation has a practical benefit as well.  The associativity of fusion products allows one to compute the fusion rules for the simple current extension $\VOA{B}$ from those of $\VOA{A}$:
\begin{align} \label{eq:IndFusion}
\Ind{\mathcal{M}} \otimes_{\VOA{B}} \Ind{\mathcal{N}}
&= \brac{\VOA{B} \otimes_{\VOA{A}} \mathcal{M}} \otimes_{\VOA{B}} \brac{\VOA{B} \otimes_{\VOA{A}} \mathcal{N}}
\cong \brac{\brac{\VOA{B} \otimes_{\VOA{A}} \mathcal{M}}\otimes_{\VOA{B}} \VOA{B}} \otimes_{\VOA{A}} \mathcal{N} \notag \\
&\cong \brac{\VOA{B} \otimes_{\VOA{A}} \mathcal{M}} \otimes_{\VOA{A}} \mathcal{N}
\cong \VOA{B} \otimes_{\VOA{A}} \brac{\mathcal{M} \otimes_{\VOA{A}} \mathcal{N}}
= \Ind{\brac{\mathcal{M} \otimes_{\VOA{A}} \mathcal{N}}}.
\end{align}
Here, we remark that inducing an irreducible $\VOA{A}$-module $\mathcal{M}$ results in an irreducible $\VOA{B}$-module $\Ind{\mathcal{M}}$.  This is a special property of (invertible) simple current extensions that follows from \eqref{eq:IndMod} because each direct summand $\SC \otimes_{\VOA{A}} \mathcal{M}$ is an irreducible that is obtained from any other direct summand by acting with one of the simple currents in $\grp{G}$, thus with an element of $\VOA{B}$.  We remark that, for simplicity and because our examples do not require it, we are ignoring the issue of \emph{fixed points} here, see \cite{FucMat96}.

Of course, one needs to justify why the fusion product of a \voa{} may be regarded as a tensor product that is permeable to the algebra action.  That fusion constitutes an abstract tensor product is, at least for $C_2$-cofinite \voas{}, one of the great results of Lepowsky's group (see \cite{HuaTen13} for a review).  While the logarithmic theories we study here are not $C_2$-cofinite in general, it is very reasonable to assume that fusion still has the properties of a tensor product on the physical category of representations.  The permeability, on the other hand, reflects the locality requirement for fields corresponding to the algebra and its modules.  More precisely, if $\func{a}{z}$ is a field of $\VOA{A}$ and $\func{\phi}{z}$ and $\func{\psi}{z}$ are fields corresponding to elements of the $\VOA{A}$-modules $\mathcal{M}$ and $\mathcal{N}$, respectively, then locality and associativity give
\begin{equation}
\begin{gathered}
\brac{\func{a}{z} \func{\phi}{x}} \func{\psi}{w} = \brac{\func{\phi}{x} \func{a}{z}} \func{\psi}{w} = \func{\phi}{x} \brac{\func{a}{z} \func{\psi}{w}} \\
\iff \qquad a \cdot \phi \otimes_{\VOA{A}} \psi = \phi \cdot a \otimes_{\VOA{A}} \psi = \phi \otimes_{\VOA{A}} a \cdot \psi.
\end{gathered}
\end{equation}

We conclude by computing the Grothendieck fusion rules of the $(1,2)$ triplet model as an application of this simple current formalism.  Recall from \eqnref{eq:TripNotMod} that the irreducible characters do not span a representation of the modular group.  From the point of view of the modular formalism of \secref{sec:ModLCFT}, one identifies candidates for the standard modules:  The typical irreducibles $\TripIrr{-1/8}$ and $\TripIrr{3/8}$ and the atypical indecomposables $\TripSt{0}$ and $\TripSt{1}$ are characterised by
\begin{equation}
\dses{\TripIrr{0}}{}{\TripSt{0}}{}{\TripIrr{1}}, \qquad \dses{\TripIrr{1}}{}{\TripSt{1}}{}{\TripIrr{0}}.
\end{equation}
Unfortunately, the standard characters do not span a representation of the modular group either.  We remark that as $\ch{\TripSt{0}} = \ch{\TripSt{1}}$, the atypical characters have measure $\tfrac{1}{3}$ with respect to the natural measure (counting measure) on the parametrisation space $M = \set{-\tfrac{1}{8}, 0, \tfrac{3}{8}}$.

As mentioned above, (Grothendieck) fusion rules for the $(1,2)$ triplet model are nevertheless easily derived by regarding this theory as the simple current extension of the $(1,2)$ singlet model.  The group $\grp{G}$ of simple currents is generated by the irreducible singlet representation $\SC = \SingIrr{2}$ (or $\SingIrr{-2}$) \cite{CreLog13} whose \hws{} has conformal dimension $3$.  We can therefore apply the above simple current formalism with $\VOA{A}$ and $\VOA{B}$ being the singlet and triplet \voas{}, respectively.  It is straightforward to identify the triplet irreducibles as induced singlet modules (see \exref{ex:Sing}):
\begin{equation}
\begin{aligned}
\TripIrr{0} &= \Ind{\SingIrr{0}}, \\
\Res{\TripIrr{0}} &= \bigoplus_{n \in \ZZ} \SingIrr{2n},
\end{aligned}
\quad
\begin{aligned}
\TripIrr{1} &= \Ind{\SingIrr{1}}, \\
\Res{\TripIrr{1}} &= \bigoplus_{n \in \ZZ} \SingIrr{2n+1},
\end{aligned}
\quad
\begin{aligned}
\TripIrr{-1/8} &= \Ind{\Fock{1/2}}, \\
\Res{\TripIrr{-1/8}} &= \bigoplus_{n \in \ZZ} \Fock{2n+1/2},
\end{aligned}
\quad
\begin{aligned}
\TripIrr{3/8} &= \Ind{\Fock{-1/2}}, \\
\Res{\TripIrr{3/8}} &= \bigoplus_{n \in \ZZ} \Fock{2n-1/2}.
\end{aligned}
\end{equation}
The remaining singlet typicals $\Fock{\lambda}$ can likewise be induced, but they turn out to define \emph{twisted} representations on which the triplet fields act with non-trivial monodromy.

The singlet fusion rules derived in \exref{ex:Sing} may now be lifted to the triplet using \eqref{eq:IndFusion}.  For example,
\begin{subequations}
\begin{equation}
\TripIrr{1} \otimes_{\VOA{B}} \TripIrr{1} = \Ind{\SingIrr{1}} \otimes_{\VOA{B}} \Ind{\SingIrr{1}} = \Ind{\brac{\SingIrr{1} \otimes_{\VOA{A}} \SingIrr{1}}} = \Ind{\SingIrr{2}} = \TripIrr{0}.
\end{equation}
In the same manner, we compute
\begin{equation}
\TripIrr{1} \otimes_{\VOA{B}} \TripIrr{-1/8} = \TripIrr{3/8}, \qquad
\TripIrr{1} \otimes_{\VOA{B}} \TripIrr{3/8} = \TripIrr{-1/8}.
\end{equation}
\end{subequations}
These results agree of course with direct, though significantly more difficult, computations \cite{GabRat96}.  For the fusion rules of $\TripIrr{-1/8}$ and $\TripIrr{3/8}$, the corresponding singlet rules were only derived at the Grothendieck level, so we may only conclude that
\begin{equation}
\ch{\TripIrr{h}} \otimes_{\VOA{B}} \ch{\TripIrr{h'}} = 2 \: \ch{\TripIrr{0}} + 2 \: \ch{\TripIrr{1}}, \qquad \text{(\(h,h' \in \set{-\tfrac{1}{8}, \tfrac{3}{8}}\)).}
\end{equation}
This result is also consistent with the known fusion rules \cite{GabRat96} where the fusion products are known to be indecomposable staggered modules formed by glueing two copies of $\TripIrr{0}$ and two copies of $\TripIrr{1}$ together.

This example illustrates how one can combine the standard module formalism of \secref{sec:ModLCFT} with the simple current formalism of \secref{sec:SimpCurrExt} to rather effortlessly deduce the (Grothendieck) fusion rules of \lcfts{}.  The results have been generalised to all the $(p',p)$ triplet models \cite{RidMod13} and are in complete agreement with the fusion computations that have been computed, as well as those that have only been conjectured, for these theories.  We conclude by remarking that we are unaware of theories to which this combined formalism cannot be applied.  It would be very interesting to study an example of such a theory.

\section*{Acknowledgements}

We thank J\"{u}rgen Fuchs and Aliosha Semikhatov for discussions on the Verlinde formula for the $(1,p)$ triplet models, Matthias Gaberdiel for helpful correspondence concerning the modular properties of the triplet characters, and Antun Milas for pointing out a mistake in an earlier version.
DR's research is supported by the Australian Research Council Discovery Project DP1093910.  
SW's work is supported by the Australian Research Council Discovery Early Career Researcher Award DE140101825.

\flushleft
\providecommand{\newblock}{}

\end{document}